\documentclass[12pt]{article}
\usepackage{graphicx}
\usepackage{appendix}
\usepackage[cm]{fullpage}
\usepackage{subcaption}
\usepackage{amsmath}
\usepackage{wrapfig}
\usepackage{CJKutf8}
\usepackage[T1]{fontenc}
\usepackage{authblk}
\usepackage{url}
\begin{document}
\begin{CJK}{UTF8}{gbsn}
\title{A Quantitative History of A.I. Research\\ in the United States and China}
\author{Daniel Ish}
\author{Andrew Lohn}
\author{Christian Curriden}
\affil{RAND Corporation, 1776 Main Street, Santa Monica, CA, USA}
\date{September 2019}

\maketitle
\begin{abstract}
Motivated by recent interest in the status and consequences of competition between the U.S. and China in A.I. research, we analyze 60 years of abstract data scraped from Scopus to explore and quantify trends in publications on A.I. topics from institutions affiliated with each country. 
We find the total volume of publications produced in both countries grows with a remarkable regularity over tens of years. 
While China initially experienced faster growth in publication volume than the U.S., growth slowed in China when it reached parity with the U.S. and the growth rates of both countries are now similar.
We also see both countries undergo a seismic shift in topic choice around 1990, and connect this to an explosion of interest in neural network methods. 
Finally, we see evidence that between 2000 and 2010, China's topic choice tended to lag that of the U.S. but that in recent decades the topic portfolios have come into closer alignment.
\end{abstract}
\pagenumbering{gobble}
\newpage
\pagenumbering{arabic}

\section{Introduction}
In July 2017, the State Council of the People's Republic of China announced its goal to make China the world leader in A.I. technology by 2030.\cite{NYT} 
This interest in developing AI technology is easily understood by considering present assessments of potential economic benefits\cite{mckaiecon} and with recent interest in the defense applications of this technology.\cite{tarraf2019department}
In the United States, this statement was received in some quarters as indicating a competitive approach to AI on the part of China and heralding a new period of technological competition.\cite{nscai}
Naturally, as with any competition, the question of who is ahead has piqued the interest of a number of observers.

A number of groups across both the US and China produce quantitative analyses of the state and recent evolution of the AI research and technology communities.\cite{allen,else,chinatech,aiind}
Generally speaking, these analyses treat publication count and impact as a proxy for the productivity of a given academic community and do tend to engage, to one extent or another, with the question of which countries have more robust AI communities. 
Some even attempt to directly answer the question of who is ahead in the AI race between U.S and China, such as recent work from the Allen Institute that takes scientific publications partitioned by impact factor as its unit of analysis.\cite{allen}
These analyses neglect any consideration of the structural features likely present in these communities, however, like the long-known dynamics of the growth of publication volume\cite{DJSP_LSBS} or the dynamics of knowledge sharing through social networks\cite{newman2001structure,hu2013hyperlinked} and across geographic\cite{usdiken1995organizational,bordersspill,jaffe1993geographic}, institutional\cite{Jaffeinst} or disciplinary\cite{diffdisc} boundaries.
Nor, indeed, do these analyses grapple with the difficult question of how research activities provide benefits to the communities in which they take place.\cite{jaffe1989real,BRANSTETTER200153,breschi2009mobility} 
Without putting analyses of the AI research communities in the US and China in the context of their internal dynamics, the relationship between the two communities and the means by which they provide value to their host communities, one cannot get a meaningful picture of who is ``ahead," the rate of change of the positions of the two communities or, indeed, the extent to which the picture of two independent entities in competition is even meaningful. 
Conversely, to the authors' knowledge, the sciencometric literature has relatively little to say about how to think about the scientific dimensions of nation-state competition.

This work is intended as a first step towards a rigorous understanding of the scientific dimensions of nation-state competition. 
Looking at long term trends in A.I. publications in the U.S. and China, we will see that both countries have exhibited remarkably stable exponential growth in the volume of publications affiliated with their institutions. 
The single exception we observe is a reduction in the rate of this exponential growth in institutions affiliated with China when their publication volume first began to match that of the U.S. between 2008 and 2009. 
We also report a metric for the overall similarity between the choice of topics in these two countries over time, and find evidence of a dramatic and persistent shift in the topic choices of both countries individually around 1990. 
Across this shift, we notice that China's choice of research topics through the 1990s and 2000s generally lagged that of the U.S., with our metric suggesting that until 2010 the choice of topics in China's research portfolio tended to more closely resemble that of the U.S. in previous years than that of the U.S. in the corresponding year. 
We offer a possible explanation for these phenomena, noting a dramatic increase of the proportion of publication on neural networks across both countries during the 1990s which then receded over time to be overtaken by a more diverse portfolio of topics. 
This shift was especially strong in China, where we note a marked overall reduction in the diversity of the topic portfolio during the 1990s. Both our similarity and diversity metrics, together with direct inspection of the popularity of selected topics, suggest that starting in 2010 these gaps closed and the modern topic portfolio of China bears close resemblance to that of the U.S.

Our results do not show any signals which map cleanly onto the onset of a competitive relationship between the two research communities. 
On the contrary, our results suggest that the A.I. research communities in these two countries appear to be closely linked and develop without being affected by the relationship between their governments. 
The apparent shift of China's publication growth when it first matched that of the U.S., the presence of the dramatic topic shift corresponding to an increase in focus on neural network methods in 1990 across both countries, and the evolution of China's topic choice to more closely match that of the U.S. all suggest that the two research communities interact and affect each other's development.
Subsequent research is needed to determine whether this phenomenon is borne out in more granular co-authorship and citation data and begin to address questions of the value each research community is providing to its host country above that which it provides to the international community.
Further research is also needed to determine the degree to which these results are specific to the AI community. 
For example, robust private sector interest in AI technology\cite{aiind} could be swamping any effect due to national priorities, which could be explored by comparing our results here with a comparable study of a technology more closely affiliated with government priorities, e.g. aerospace research and development.
Beyond explorations of the role of science and technology in international competition, the metrics presented here form a lightweight suite of techniques for exploring the relationship of research communities across time and boundaries that is complimentary with studies of citation and co-authorship.

\section{The Data}

Defining the boundaries of the field of artificial intelligence can be a research question unto itself.\cite{else} Typically, AI is defined as the set of tasks technologies which allow computers to complete tasks which seem to require intelligence, most frequently though not exclusively through the use of machine learning, a set of techniques for using computers to discover statistical relationships in data.\cite{NAP25488} This includes tasks like image and speech recognition and natural language processing.\cite{else} This focus on machine learning techniques is a modern development in the field, with older research utilizing a wide variety of techniques, including formal logic.\cite{buchanan2005very}

The subject of our study is data scraped from the web interface of Scopus, an abstract and citation database run by Elsevier. 
We automatically queried Scopus for articles affiliated with China or the U.S. with certain A.I. related keywords in the title, abstract or keyword fields and recorded the number of results by year. Scopus defines article country affiliation as the country in which the author's affiliated institution is located. 
Keywords were generated from the methodologies described in the scikit-learn documentation as of January 2019 and topics listed in the tutorials and workshops at the 2018 NeurIPS, ICML, and AAAI conferences. 
Scikit-learn was the second most popular open-source machine learning software in 2019 as measured by cumulative GitHub "stars"\cite{aiind} and contains extensive documentation describing the wide variety of techniques supported.
NeurIPS, ICML and AAAI were the largest, third largest and fifth largest AI conferences by attendance in 2019, with NeurIPS and ICML exhibiting the fastest growth.\cite{aiind}
We included terms from these resources among our search keywords if, in the judgment of the authors, they represented a topic or method specific to AI research.
We kept results starting in 1981 for papers affiliated with China, since 1980 was the latest year for which we did not find any publications associated with our keywords and affiliated with China on Scopus. 
For the U.S., we kept results starting in 1962, as this was the earliest year for which we found data. 
For both countries, we only analyzed data up to 2018, since that was the last full year of data when the analysis was conducted.

Before proceeding with our analysis, we offer a few caveats about the quality of the data we obtained this way. The first and most glaring issue is that our data collection methodology does not capture any information about publications that are not written in English. As such, our estimate of publication volume affiliated with China are probably a significant underestimate and it is possible that the overall topic portfolio of research affiliated with China is different from what we see in our data. Indeed, while our search for ``machine learning" returned $4{,}365$ articles affiliated with China on Scopus, a search for the Mandarin equivalent (``机器学习''） on CNKI, a repository of scholarly publications Chinese scholarly publications, returned $10{,}733$ results,\footnote{On 8/21/2019} suggesting rough parity between the size of the English and Mandarin literatures. In their recent work tracking China's overall scientific output, Qingnan Xie and Richard Freeman found this to be true of scientific articles in general.\cite{bigger} We did not collect information on citations for each paper, so we have no proxy for the quality or impact of a particular paper. Our analysis also relies on assigning national affiliation by institution, which elides the tricky question of how to assign the research production of, for example, a Chinese national affiliated with a U.S. university or a U.S. national affiliated with a Chinese A.I. research firm. While this omission seems especially acute in our dataset, it is emblematic of the difficulty inherent in analyzing entangled international institutions like research communities in the frame of competition between nation states.

We should also point out that the way we assembled our dataset and chose our keywords is somewhat ad hoc and biased towards the modern conception of the field in the U.S. technology sector. One might reasonably be concerned about missing research on substantially similar topics, particularly the farther into the past one looks, possibly even connected to the citation network of some of the papers we do assemble. Based on some checks performed by hand, it is also likely that some number of publications which the authors would not consider artificial intelligence or machine learning papers made it into the dataset. For example, we observed psychology papers among the results for our search of ``transfer learning." We are somewhat concerned about the implications for our results on the keyword ``logistic regression'' in particular, since a large number of other fields of science use logistic regression as method for statistical inference. Ultimately, this presents a definitional problem like that tackled in the Elsevier study\cite{else} and we proceed under the theory that further manual tweaks to our dataset to try to address these deficiencies are as likely to introduce additional bias as remove any. For this work, we leave it to the reader to decide how substantively they believe these considerations affect the results presented and look forward to future work which explores this topic with a more robust data collection process.

\section{Total Volume and Lag}

As a simplifying assumption to enable our analysis, we take each of these keywords to be mutually exclusive. That is, we assume that the papers returned in the search for each keyword do not appear in the searches for any other keyword. Put another way, we ignore any double counting that has occurred in the process of assembling our data. This is certainly not true in reality (e.g. machine vision papers are likely to utilize neural networks), but is hopefully close enough to true as a first approximation to provide some insight. To address the question of trends in Chinese and U.S. publication volume relative to one another, we develop two metrics. The first of these is simply the total number of publications we observed under the assumption of mutual exclusivity. That is, if $n_C(y,k)$ (resp. $n_U(y,k)$) is the total number of publications observed with keyword $k$ in year $y$ from China (resp. the U.S.), we track
\begin{align}
n_I(y) = \sum_k n_I(y,k)
\end{align}
as our proxy for the total number of papers produced, where $I= C$ or $U$. A visual inspection of the data suggests that the total volume for the U.S. is exponential over the whole period of study, so we fit it to the model
\begin{align}
\ln(n_U(y)) = m_U y + b_U + \epsilon_U(y)
\end{align}

\begin{wrapfigure}{r}{0.5\textwidth}
\includegraphics[page=1]{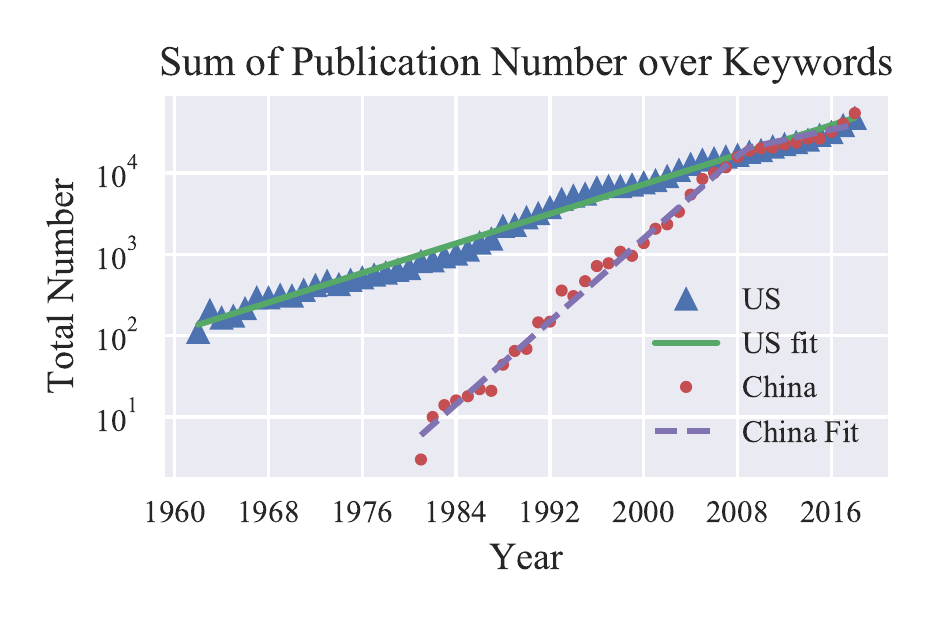}
\includegraphics[page=2]{all_plots.pdf}\\
\centering
\begin{tabular}{l||r|r|r|}
& Fit Value & Std. Err. & $t$-statistic\\
\hline\hline
$m_U$ & 0.1047 & 0.001 & 47.568\\
$m_C$ & 0.2935 & 0.004 & 66.718\\
$\mu_C$ &  -0.2210 & 0.015 & -14.756\\
$m_\delta$ & -0.5964 & 0.008 & -76.245
\end{tabular}
\caption{\label{volume}Plots and linear fits of keyword-averaged lag and total volume. Note the logarithmic scale on the plots of total volume.}
\end{wrapfigure}

For China, the data also appears to follow an exponential growth pattern, though there is some suggestion that the rate of growth of the exponential changes when the publication volume of China first matches that of the U.S. between 2008 and 2009. So, we fit the total volume data for China to the model
\begin{align}
\ln(n_C(y)) = m_C y + \mu_C(y - y_0)_+ +b_C + \epsilon_C(y)
\end{align}
where
\begin{align}
(x)_+ = \left\{\begin{array}{lr}x & x\geq 0\\0 & x<0\end{array}\right.
\end{align}
\noindent and $y_0$ is chosen to be the point in time where we estimate that Chinese total publication volume first met that of the U.S. by linear interpolation between the data points in 2008 and 2009. $\mu_C$ is the magnitude of the change in slope of the log-linear relationship, i.e. the putative change in the rate of growth of the exponential. 

The second metric is somewhat more circuitous, but does not depend on our assumption of mutual exclusivity. We define the ``lag" of the keyword $k$ at year $y$, $\delta_k(y)$, to be the smallest integer such that
\begin{align}
n_C(y+\delta_k(y),k) \geq n_U(y,k)
\end{align}
if such an integer exists. That is, the lag is the number of years by which Chinese publication volume is behind the U.S. publication volume in keyword $k$. For example, in 1987, the U.S. volume of papers with the keyword ``neural network" was $75$, and China did not meet or exceed that volume of papers with the keyword ``neural network" until 1993 when we observed $220$ such papers. Thus the lag for the year 1987 and keyword ``neural network" was 6 years. We then calculate the lag averaged over keywords
\begin{align}
\overline{\delta}(y) = \frac{1}{m(y)}\sum_k \delta_k(y)
\end{align}
where $m(y)$ is the number of keywords in year $y$ for which $\delta_k(y)$ exists, and fit it to the model
\begin{align}
\overline{\delta}(y) = m_\delta y + b_\delta + \varepsilon(y)
\end{align} 

The relationship between total volume and lag is somewhat complicated. If every keyword were individually exponential in time with the same rate of growth and $\mu_C = 0$ (which is not the case in our data), we would expect that
\begin{align}\label{Check}
m_\delta = \frac{m_U - m_C}{m_C}
\end{align}
That is, if all the keywords grew at the same rate the slope of the lag would be related to the rates of growth of the total volumes by Equation~\ref{Check}. However, this relationship is no longer guaranteed under weaker assumptions. To see this, note that the lag weights each keyword individually while the total volume fit weights them by their observed volume. That is, this relationship could be violated if our data had a large number of low-volume keywords and a few high volume keywords with radically different time dependences. Put another way, the lag incorporates some information about how the keyword distributions compare and some information about the volume over time within each keyword and, generally speaking, rewards diverse research portfolios. The lag would remain large, for example, if China's growth to match the U.S. in total publication volume occurred only within a few keywords with the U.S. continuing to lead by large numbers of years in many other keywords. Furthermore, it is mathematically possible for the lag to not assume a linear dependence on time despite the total volumes from each country remaining exponential in time.

We obtain the fit parameters for all these models with the ordinary least squares estimator, as this estimator remains consistent under any structure for the covariance matrix of the errors. The Durbin-Watson statistic of each of these fits suggest positive autocorrelation, however, so to quantify the extent to which these fit parameters differ from zero we use a Bartlett kernel HAC estimator with bandwidth $\sqrt{N}$, where $N$ is the number of data points, to calculate the standard error of each of the fit parameters.\cite{sun2008optimal} Thus, the ``$t$-statistic" reported does not have a $t$ distribution in finite samples, though it remains asymptotically normally distributed.

In general, these fits are quite convincing. The $t$ statistics for the parameters of interest listed in the table in Figure~\ref{volume} are all quite large, assuaging any anxiety we might have about the finite sample distributions of these quantities. This analysis suggests that publication volume on these topics in both countries grew exponentially in time, and that the time constant of China's growth shifted to come into more close alignment with that of the U.S. when the two countries began to have similar publication volumes. This could be due, for example, to the growth of both countries' publications being limited by a third independent process, e.g. the growth in the number of academic journals accepting publications on these topics. The right-hand quantity in Equation~\ref{Check} is more negative than $m_\delta$ by more than $5$ times the standard error in $m_\delta$. This is easily explained by our subsequent explorations of the keyword distribution, as China's heavy focus on neural networks relative to the U.S. during its period of initial growth likely resulted in a larger lag value than simply observing the volume would suggest. This results in the lag suggesting that the moment China ``caught up" to the U.S. was around mid-2011, later than the date suggested by the total volume, in mid-2008. It is remarkable that the dynamics of the diversity of each country's portfolio interact with those of the total volume to produce a linear dependence of the lag on time, despite the complicated structure of topic choice over time we will see below.

\pagebreak

\section{Topic Distributions}

\begin{wrapfigure}{r}{0.5\textwidth}
\includegraphics[page=4]{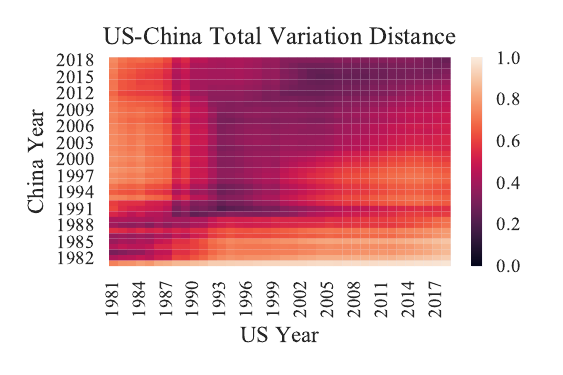}
\includegraphics[page=5]{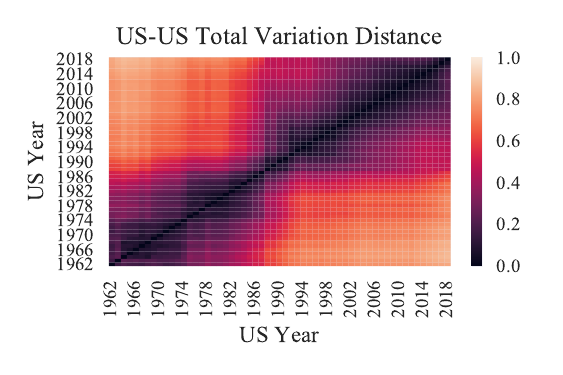}
\includegraphics[page=6]{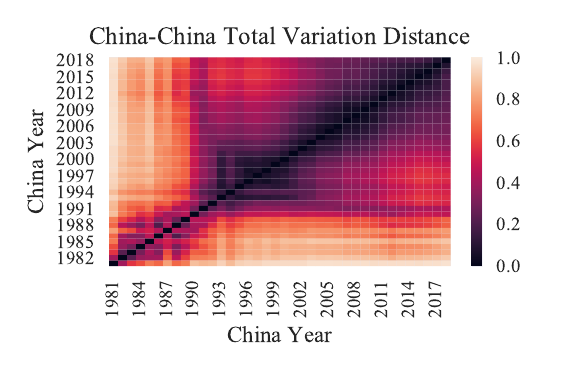}
\caption{\label{TVD}Plots of the total variation distance across both country and year. Darker colors indicate more similar topic choice.}
\end{wrapfigure}

We turn now to attempting to quantify and explore the similarities of topic choice both between countries and over time. We interpret the quantity
\begin{align}\label{prob}
\hat{p}_I(y,k) = \frac{n_I(y,k)}{n_I(y)}
\end{align}
as being an estimate of the underlying probability for a paper produced by country $I$ in year $y$ to possess keyword $k$. We then investigate the total variation distance
\begin{align}
\Delta_{I,I^\prime}(y,y^\prime) = \frac{1}{2}\sum_k \left|\hat{p}_{I}(y,k) - \hat{p}_{I^\prime}(y^\prime,k)\right|
\end{align}
as our measure of the difference between the distribution of keywords of country $I$ in year $y$ and that in country $I^\prime$ in year $y^\prime$. Under the mutual exclusivity assumption, this quantity is the maximum difference in probability the two distributions would assign to an event.

Even absent this assumption, however, we can still regard $\hat{p}_I(y,k)$ for fixed $I$ and $y$ as a random vector of unit $\ell_1$ norm which captures some information about the distribution of topics. $\Delta$ is then simply half the $\ell_1$ distance between these vectors, which still captures some idea of the distance between the topic distribution in different years and countries. In either interpretation, smaller values of $\Delta$ indicate that the topic portfolios being compared bear a closer relationship to one another.

Looking to Figure~\ref{TVD}, we find a remarkable amount of structure. Both counties show a marked transformation of their research portfolio around 1990 that persists to the present day. This is perhaps most striking in the case of the U.S., which has significant autocorrelation both before and after this sudden shift. We see smaller shifts take place both before and after this realignment in the U.S. portfolio, suggesting a balance of topics slowly shifting over time. China does not show the same clear, marked autocorrelation before about 1990, but this is easily understood by returning to Figure~\ref{volume} and noticing that before about 1990 China was producing less that 100 English language publications on these topics per year. The quantity in Equation~\ref{prob} will be more noisy for smaller sample sizes, and so too will our estimate of the total variation distance between the underlying probabilities.

\begin{figure}[t!]
\vspace{1pt}
\includegraphics[page=8]{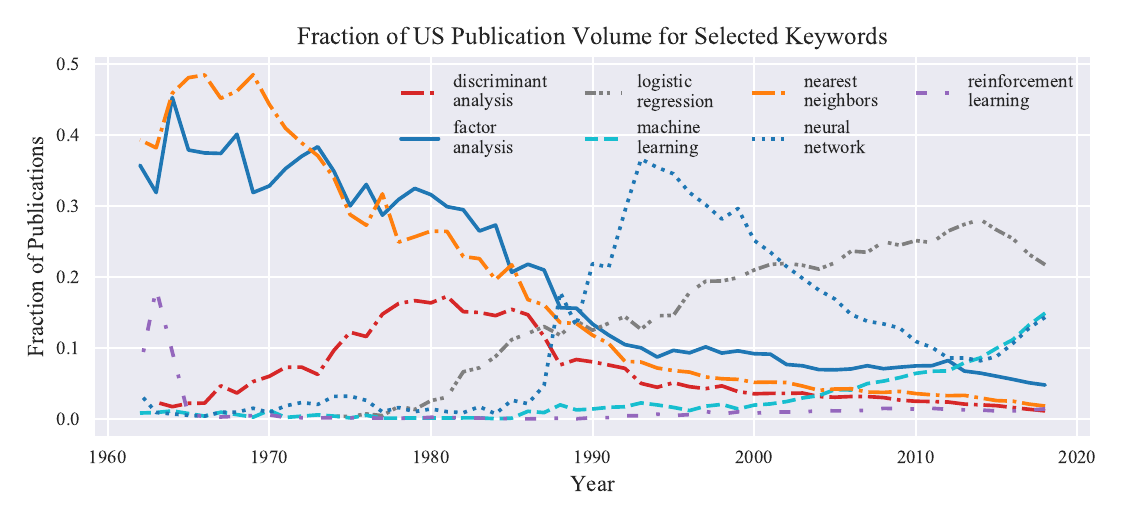}
\includegraphics[page=9]{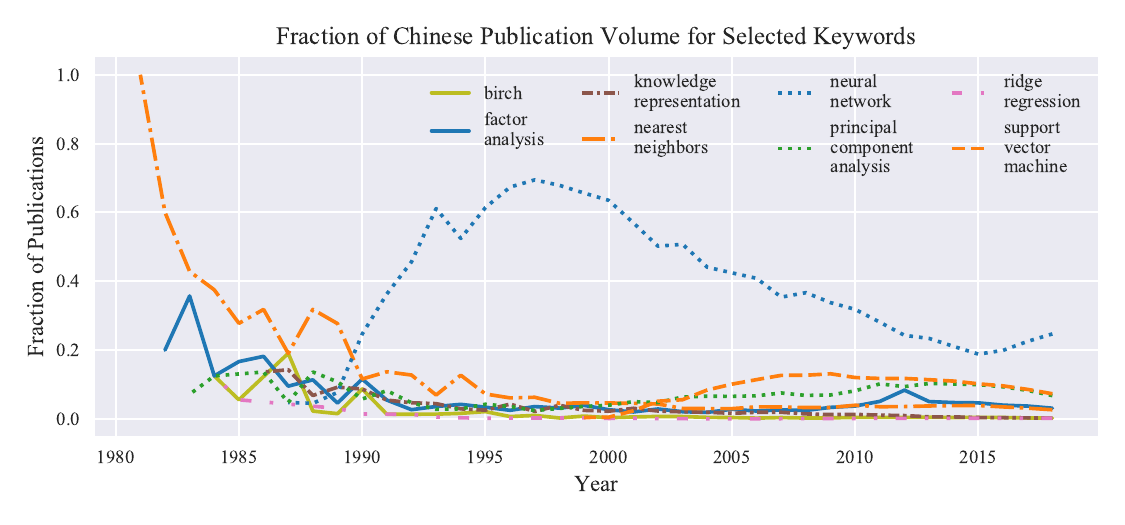}
\caption{\label{time} Plots of the fraction of publication volume for selected keywords over time. Keywords were included if they contributed at least 12\% of total volume from the country in question in at least one year.}
\end{figure}

Considering the third plot in Figure~\ref{TVD}, which depicts the total variation distance between the keyword prevalence in the U.S. and China across all available years, we see some evidence of similarity between the research programs of the two countries which carries across the large shift around 1990. Before 1990, this relationship is more noisy, likely due to the low volume of English language publications out of China during that time period. The two countries had similar research programs during the 1990s, during which time both exhibited visible autocorrelation. During the 2000s, however, China's research program bore more resemblance to the research program of the U.S. in the 1990s than that of the 2000s. In the past decade, however, China's research program has become more closely aligned with the U.S. program from a broad base of years between 2000 and the present, mirroring the autocorrelation of the current U.S. research program.

\begin{figure}[t]
\vspace{1pt}
\includegraphics[page=7]{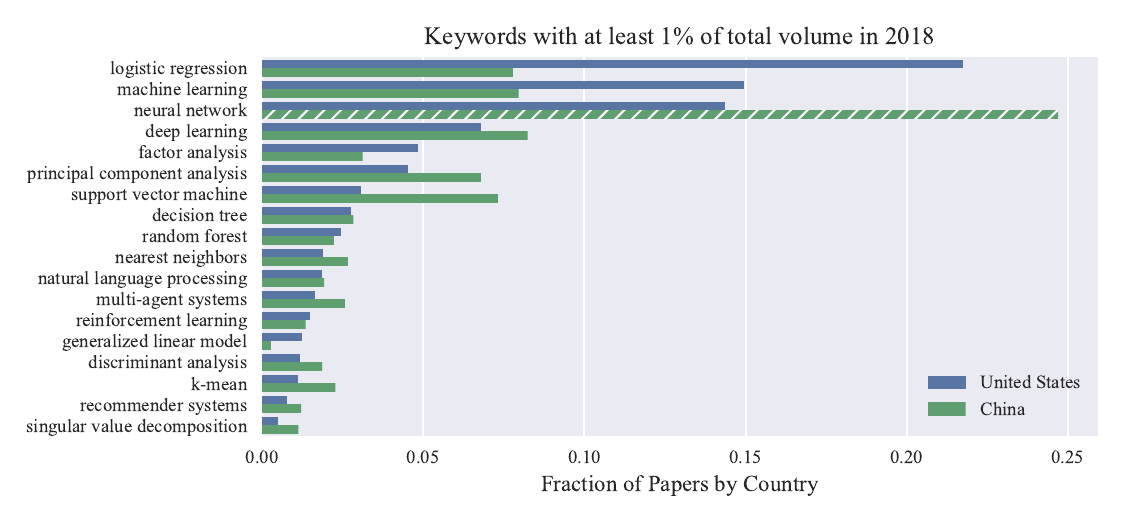}
\caption{\label{bar} Fraction of total publication volume in 2018 by country for keywords which make up at least 1\% of total volume of at least one country in that year.}
\end{figure}

\begin{wrapfigure}{r}{0.5\textwidth}
\includegraphics[page=3]{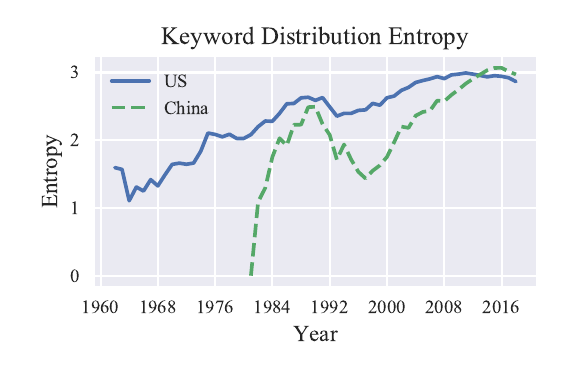}
\caption{\label{ent} The entropy of the two publication distributions (as estimated by Equation~\ref{enteq}) over time.}
\end{wrapfigure}

While the total variation distance provides an attractive measure of the similarity between the research programs of different countries and years which captures the contributions of all keywords, it does not easily expose which keywords actually participated in these shifts in research nor admit a clear intuition for how similar two research programs actually are given their total variation difference. Figure~\ref{time} addresses the first question, allowing us in particular to assign the seismic shift that occurred in research focus around 1990 as due at least in part to the rise of neural networks. Similarly, we can see one possible explanation for the phenomenon observed in the third plot of Figure~\ref{TVD} in that the proportion of U.S. publication volume devoted to neural networks waned much more quickly at the turn of the millennium than that of Chinese publication volume.  

Though the less popular topics which still met the criterion for inclusion in Figure~\ref{time} are different between the two countries, we can look to Figure~\ref{bar} to see that the leading topics in both countries tend to appear in similar though distinct proportions, at least in 2018. The two research programs are different in emphasis rather than wholly different, at least by this analysis. This presents us with a bit more context on why the waning of popularity of neural networks in China began to bring China back into closer alignment with the U.S.: the next most popular topics are, broadly speaking, shared between both countries. 

Looking at Figure~\ref{time}, we see some evidence that though the U.S. shifted some focus onto neural networks, China shifted its focus much more aggressively. Some of that focus remains even today, as we can see in Figure~\ref{bar}, which still shows neural networks making up a greater fraction of China's publication than any single topic makes up of the U.S. publications. In order to probe the extent to which this focus impacts the overall diversity of the research portfolio in China, we study the entropy
\begin{align}
S_I(y) = - \sum_k \hat{p}_I(y,k)\ln \hat{p}_I(y,k)\label{enteq}
\end{align}
of the publication distribution as a probe of the overall diversity of the research portfolio. If a country were to publish all of its on a single topic in a year, we would have $S_I(y)=0$. With our $64$ keywords, the maximum value $S_I(y)$ can take is $\ln(64)\approx 4.16$, which would indicate a country that publishes equally on all topics.

Figure~\ref{ent} confirms quantitatively one of our suspicions from looking at Figure~\ref{time}. During the 1990s, the overall topic diversity of China's English language publication output declined precipitously due to their focus on neural networks. Conversely, though the U.S. entropy shows a slight decline in the 1990s, the rise of neural networks did not affect the overall diversity of U.S. publication nearly as strongly. Notably, as neural networks waned in popularity in China we see the overall diversity of publication topics rise to eventually exceed that of the U.S.

\section{Conclusions}

Though the U.S. and China reacted to different degrees to the recent interest in neural network methods, the data shows a dramatic consonance between the research programs in both time and topic choice. China was and remains more focused on neural network methods than the U.S., at least by this data, but both countries clearly explore the same areas and react to the same trends in research topics. Notably, most of the regions of lower total variation distance between the two countries shown in Figure~\ref{TVD} occur above the diagonal. That is, from about 1990 through to the present, the portfolio of topics that institutions in China produced publications on in any given year bore more resemblance to the U.S. in previous years than it did to the U.S. in that year or future years. This is consistent with a picture wherein China lagged the U.S. not just in volume but also in topic choice. Without further study, however, it's difficult to say whether this indicated that China's research program was actually any less effective during that time.

Further study is also needed to determine how a competitive posture between two governments affects the interaction of their scientific communities, and indeed if it does at all. 
If some effect is found, to what degree is it heterogeneous across disciplines? 
Most critically, does this effect actually correspond to each community providing benefits to its host nation above those that it provides to the international community as a whole?
Though we have plenty of historical and anecdotal reason to suspect that science plays an important role in nation-state competition, at present there is no rigorous, quantitative accounting of this role.
\bibliography{ref}

\begin{thebibliography}{10}

\bibitem{NYT}
Paul Mozur.
\newblock Beijing wants {A.I.} to be made in {China} by 2030.
\newblock {\em New York Times}, 20, 2017.
\newblock
  \url{https://www.nytimes.com/2017/07/20/business/china-artificial-intelligence.html}
  (Accessed 8/23/2019).

\bibitem{mckaiecon}
Jacques Bughin, Jeongmin Seong, James Manyika, Michael Chui, and Raoul Joshi.
\newblock Notes from the ai frontier: Modeling the impact of ai on the world
  economy.
\newblock {\em McKinsey Global Institute}, 2018.

\bibitem{tarraf2019department}
Danielle~C. Tarraf, William Shelton, Edward Parker, Brien Alkire,
  Diana~Gehlhaus Carew, Justin Grana, Alexis Levedahl, Jasmin Leveille, Jared
  Mondschein, James Ryseff, Ali Wyne, Dan Elinoff, Edward Geist, Benjamin~N.
  Harris, Eric Hui, Cedric Kenney, Sydne Newberry, Chandler Sachs, Peter
  Schirmer, Danielle Schlang, Victoria~M. Smith, Abbie Tingstad, Padmaja
  Vedula, and Kristin Warren.
\newblock The department of defense posture for artificial intelligence.
\newblock {\em RAND Corporation}, 2019.
\newblock \url{https://www.rand.org/pubs/research_reports/RR4229.html}.

\bibitem{nscai}
National Security~Commission on~Artificial~Intelligence.
\newblock Interim report.
\newblock 2019.
\newblock \url{https://www.nscai.gov/reports}.

\bibitem{allen}
Field Cady and Oren Etzioni.
\newblock China may overtake {US} in {AI} research.
\newblock Allen Institute for Artificial Intelligence,
  \url{https://medium.com/ai2-blog/china-to-overtake-us-in-ai-research-8b6b1fe30595}
  (Accessed 8/23/2019), 2019.

\bibitem{else}
Artificial intelligence: How knowledge is created, transferred, and used.
\newblock Technical report, Elsevier, 2018.

\bibitem{chinatech}
Global artificial intelligence industry data report
  [全球人工智能产业数据报告].
\newblock Technical report, China Academy of Information and Communications
  Technology Data Research Center
  [中国信息通信研究院数据研究中心], 2019.
\newblock
  \url{http://www.caict.ac.cn/kxyj/qwfb/qwsj/201905/P020190523542892859794.pdf}(Accessed
  8/22/19), Translation by Joy Dantong Ma and Jeffrey Ding available at
  \url{https://chinai.substack.com/}, (Accessed 8/22/19), see issue \#62.

\bibitem{aiind}
Raymond Perrault, Yoav Shoham, Erik Brynjolfsson, Jack Clark, John Etchemendy,
  Barbara Grosz, Terah Lyons, James Manyika, Saurabh Mishra, and Juan~Carlos
  Niebles.
\newblock The ai index 2019 annual report.
\newblock {\em AI Index Steering Committee, Human-Centered AI Institute,
  Stanford University, Stanford, CA}, 2019.

\bibitem{DJSP_LSBS}
Derek~John de~Solla~Price.
\newblock {\em Little science, big science}.
\newblock Columbia University Press New York, 1965.

\bibitem{newman2001structure}
Mark~EJ Newman.
\newblock The structure of scientific collaboration networks.
\newblock {\em Proceedings of the national academy of sciences},
  98(2):404--409, 2001.

\bibitem{hu2013hyperlinked}
Yansong Hu.
\newblock Hyperlinked actors in the global knowledge communities and diffusion
  of innovation tools in nascent industrial field.
\newblock {\em Technovation}, 33(2-3):38--49, 2013.

\bibitem{usdiken1995organizational}
Behl{\"u}l {\"U}sdiken and Yorgo Pasadeos.
\newblock Organizational analysis in north america and europe: A comparison of
  co-citation networks.
\newblock {\em Organization studies}, 16(3):503--526, 1995.

\bibitem{bordersspill}
Jasjit Singh and Matt Marx.
\newblock Geographic constraints on knowledge spillovers: Political borders vs.
  spatial proximity.
\newblock {\em Management Science}, 59(9):2056--2078, 2013.

\bibitem{jaffe1993geographic}
Adam~B Jaffe, Manuel Trajtenberg, and Rebecca Henderson.
\newblock Geographic localization of knowledge spillovers as evidenced by
  patent citations.
\newblock {\em the Quarterly journal of Economics}, 108(3):577--598, 1993.

\bibitem{Jaffeinst}
Adam~B. Jaffe and Manuel Trajtenberg.
\newblock Flows of knowledge from universities and federal laboratories:
  Modeling the flow of patent citations over time and across institutional and
  geographic boundaries.
\newblock {\em Proceedings of the National Academy of Sciences},
  93(23):12671--12677, 1996.

\bibitem{diffdisc}
Erjia Yan.
\newblock Disciplinary knowledge production and diffusion in science.
\newblock {\em Journal of the Association for Information Science and
  Technology}, 67(9):2223--2245, 2016.

\bibitem{jaffe1989real}
Adam~B Jaffe.
\newblock Real effects of academic research.
\newblock {\em The American economic review}, pages 957--970, 1989.

\bibitem{BRANSTETTER200153}
Lee~G Branstetter.
\newblock Are knowledge spillovers international or intranational in scope?:
  Microeconometric evidence from the u.s. and japan.
\newblock {\em Journal of International Economics}, 53(1):53 -- 79, 2001.

\bibitem{breschi2009mobility}
Stefano Breschi and Francesco Lissoni.
\newblock Mobility of skilled workers and co-invention networks: an anatomy of
  localized knowledge flows.
\newblock {\em Journal of economic geography}, 9(4):439--468, 2009.

\bibitem{NAP25488}
{National Academies of Sciences, Engineering, and Medicine}.
\newblock {\em Implications of Artificial Intelligence for Cybersecurity:
  Proceedings of a Workshop}.
\newblock The National Academies Press, Washington, DC, 2019.

\bibitem{buchanan2005very}
Bruce~G Buchanan.
\newblock A (very) brief history of artificial intelligence.
\newblock {\em Ai Magazine}, 26(4):53--53, 2005.

\bibitem{bigger}
Qingnan Xie and Richard~B Freeman.
\newblock Bigger than you thought: China's contribution to scientific
  publications and its impact on the global economy.
\newblock {\em China \& World Economy}, 27(1):1--27, 2019.

\bibitem{sun2008optimal}
Yixiao Sun, Peter~CB Phillips, and Sainan Jin.
\newblock Optimal bandwidth selection in heteroskedasticity--autocorrelation
  robust testing.
\newblock {\em Econometrica}, 76(1):175--194, 2008.

\end{thebibliography}

\appendix
\section{Full List of Search Terms}
\begin{tabular}{ccc}
graph-based learning & text mining & reinforcement learning\\[1ex]
imitation learning & deep learning & relational learning\\[1ex]
heuristic search & knowledge representation & machine translation\\[1ex]
machine learning & multi-agent systems & policy learning\\[1ex]
generative model & bayesian methods & meta-learning\\[1ex]
machine vision & recommender systems & causal learning\\[1ex]
multi-task learning & causal inference & natural language processing\\[1ex]
transfer learning & generative adversarial network & cognitive systems\\[1ex]
bayesian regression & polynomial regression & logistic regression\\[1ex]
discriminant analysis & least angle regression & decision tree\\[1ex]
multi-layer perceptron & nearest neighbors & latent semantic analysis\\[1ex]
latent dirichlet allocation & support vector machine & random forest\\[1ex]
singular value decomposition & ordinary least squares & spectral clustering\\[1ex]
adaboost & agglomerative clustering & k-mean\\[1ex]
dictionary learning & isotonic regression & elastic net\\[1ex]
lasso & gaussian mixture & independent component analysis\\[1ex]
principal component analysis & perceptron & neural network\\[1ex]
ridge regression & mean-shift & kernel ridge regression\\[1ex]
generalized linear model & factor analysis & gaussian process regression\\[1ex]
naive bayes & affinity propagation & dbscan\\[1ex]
orthogonal matching pursuit & non-negative matrix factorization & stochastic gradient descent\\[1ex]
birch &  & \\[1ex]
\end{tabular}

\end{CJK}
\end{document}